\documentclass[preprint,aps]{revtex4}

\usepackage{graphicx}
\usepackage{dcolumn}
\usepackage{bm}
\usepackage{epsf}

\newcommand{\beq}{\begin{equation}}
\newcommand{\eeq}{\end{equation}}
\newcommand\beqa{\begin{eqnarray}}
\newcommand\eeqa{\end{eqnarray}}
\newcommand\bea{\begin{array}}
\newcommand\eea{\end{array}}
\newcommand{\nn}{\nonumber}
\newcommand{\neqa}{\nonumber\end{eqnarray}}
\newcommand{\la}{\label}

\newcommand{\Eq}[1]{Eq.(\ref{#1})}
\newcommand{\Eqs}[2]{Eqs.(\ref{#1},\ref{#2})}
\newcommand{\ur}[1]{(\ref{#1})}
\newcommand{\urs}[2]{(\ref{#1},\ref{#2})}

\newcommand{\ee}{\epsilon}

\newcommand{\zh}[1]{{\bf {#1}}}

\def\Dirac#1{#1\hskip-6pt/}
\def\dd{\Dirac\partial}

\begin{document}

\title{Quark structure of chiral solitons}

\vspace{1cm}

\author{\bf Dmitri Diakonov}

\vspace{1cm}

\affiliation{Thomas Jefferson National Accelerator Facility, Newport News, VA 23606, USA\\
NORDITA, Blegdamsvej 17, DK-2100 Copenhagen, Denmark\\
St.~Petersburg Nuclear Physics Institute, Gatchina, 188300, St.~Petersburg, Russia}

\date{August 16, 2004}

\vspace{1cm}

\begin{abstract}
There is a prejudice that the chiral soliton model of baryons is something
orthogonal to the good old constituent quark models. In fact, it is the opposite:
the spontaneous chiral symmetry breaking in strong interactions explains the
appearance of massive constituent quarks of small size thus justifying the constituent
quark models, in the first place. Chiral symmetry ensures that constituent quarks
interact very strongly with the pseudoscalar fields. The ``chiral soliton'' is another
word for the chiral field binding constituent quarks. We show how the old $SU(6)$
quark wave functions follow from the ``soliton'', however, with computable
relativistic corrections and additional quark-antiquark pairs. We also find the
5-quark wave function of the exotic baryon~$\Theta^+$~\cite{F1}.
\end{abstract}

\maketitle

\section{The necessity of quantum field theory}

It has been known since the work of Landau and Peierls (1931) that the 
quantum-mechanical wave function description, be it non-relativistic or relativistic, 
fails at the distances of the order of the Compton wave length of the particle.
Measuring the electron position with an accuracy better than $10^{-11}\,{\rm cm}$ 
produces a new electron-positron pair, by the uncertainty principle. One observes
it in the Lamb shift and other radiative corrections. Fortunately, the atom 
size is $10^{-8}\,{\rm cm}$, therefore there is a gap of three orders of magnitude 
where we can successfully apply the Dirac or even the Schr\"odinger equation. 
In baryons, we do not have this luxury. Measuring the quark position with an accuracy 
higher than the {\em pion} Compton wave length of $1.4\,{\rm fm}$ produces a pion, 
{\it i.e.} a new quark-antiquark ($Q\bar Q$) pair, whereas the baryon size is $0.8\,{\rm fm}$. 
Therefore, there seems to be no room for the quantum-mechanical wave function description 
of baryons at all. To describe baryons, one needs a quantum field theory from the start, 
with a varying number of $Q\bar Q$ pairs, because of the spontaneous chiral symmetry 
breaking which makes pions light. 

Ignoring quantum field theory where it cannot be ignored, causes multiple problems. 
Let me mention just two paradoxes of the standard constituent quark models, out of many.

The first is the value of the so-called nucleon sigma term~\cite{Arndt1}. 
It is experimentally measured in low-energy $\pi N$ scattering, 
and its definition is the scalar quark density in the nucleon, multiplied 
by the current (or bare) quark masses,
$$\sigma=\frac{m_u+m_d}{2}\,<N|\bar u u+\bar d d|N>=67\pm 6\,{\rm MeV}.$$
The standard values of the current quark masses are  
$m_u\simeq 4\,{\rm MeV},\,m_d\simeq 7\,{\rm MeV}$ (and $m_s\simeq 150\,{\rm MeV}$).
In the non-relativistic limit, the scalar density is the same as the
vector density; therefore, in this limit the matrix element above is
just the number of $u,d$ quarks in the nucleon, equal to 3. If $u,d$
quarks are relativistic, the matrix element is strictly less than three. 
Hence, in the naive constituent quark model
$$\sigma_{\rm quarks}\leq \frac{4\,{\rm MeV}+7\,{\rm MeV}}{2}*3
=17.5\,{\rm MeV},$$
that is {\it four} times less than experimentally! {\it Three quarters} of the
$\sigma$ term is actually residing not in the three constituent quarks but in
the additional quark-antiquark pairs in the nucleon.

The second paradox which is probably less known, arises when one attempts to
extract quark distributions as function of Bjorken $x$ from a constituent quark
model, be it any variant of the bag model or any variant of the potential models
with any kind of correlations between quarks. If the three quarks are loosely bound,
their distribution function is just $\delta\left(x-\frac{1}{3}\right)$, each quark
carrying $1/3$ of the nucleon momentum in the infinite momentum frame. As quarks
become more bound, this $\delta$-function is smeared around $1/3$. 
However, higher quark velocities imply that the ``lower'' component of the Dirac bispinor  
wave function increases (it is zero in the extreme non-relativistic case), 
at the expense of the decrease of the ``upper'' component. It means that if quarks 
are moving inside a nucleon, there are {\em less} than three quarks in the nucleon. 
Since the number of quarks {\em minus} the number of antiquarks is the conserved 
baryon number, it automatically means that the number of antiquarks is {\em negative}~\cite{SF}. 
It is an inevitable mathematical consequence of the Dirac equation. The paradox 
is cured by adding the Dirac sea to valence quarks; only then the antiquark 
distribution becomes positive-definite, and satisfies the general sum rules~\cite{SF}. 

Thus, a field-theoretic description of baryons is a must if one does not 
wish to violate general theorems, and also for practical reasons.  

I present below a relativistic field-theoretic model of baryons where the
above paradoxes are resolved, together with the well-known ``spin crisis''
paradox. Actually, one has to be surprised not by why the constituent
quark approach is a failure but rather why does it work at all in a variety
of cases. The model will answer this question, too.

\section{The chiral quark -- soliton model}

The most important happening in QCD from the point of view of the light hadron
structure is the Spontaneous Chiral Symmetry Breaking (SCSB): as its result, almost 
massless $u,d,s$ quarks get the dynamical momentum-dependent masses $M_{u,d,s}(p)$, 
and the pseudoscalar mesons $\pi, K, \eta$ become light (pseudo) Goldstone bosons.
At the same time, pseudoscalar mesons are themselves bound $Q\bar Q$ states.
How to present this queer situation mathematically? There is actually not much freedom
here: the interaction of pseudoscalar mesons with constituent quarks is dictated
by chiral symmetry. It can be written in the following compact form~\cite{DP86}:
\beq 
L_{\rm eff}=\bar q\;\left[i\dd-M\exp(i\,\gamma_5\,
\pi^A \lambda^A/F_\pi) \right]\,q,\quad \pi^A=\pi,K,\eta. 
\la{lagrangian}\eeq 
Since \Eq{lagrangian} is an effective low-energy theory, one expects formfactors 
in the constituent quark -- pion interaction; in particular, $M(p)$ is 
momentum-dependent~\cite{DP86} and provides an UV cutoff. In fact, \Eq{lagrangian} 
is written in the limit of zero momenta. A possible wave-function renormalization 
factor $Z(p)$ can be also admitted but it can be absorbed into the definition of the 
quark field. Notice, that there is no kinetic-energy term for pseudoscalar fields
in \Eq{lagrangian}. It is in accordance with the fact that pions are not ``elementary'' 
but a composite field, made of constituent quarks. The kinetic energy term (and all 
higher derivatives) for pions appears from integrating out quarks, or, in other words, 
from quark loops, see Fig.~1. 

\begin{figure}[b]
\centerline{\epsfxsize=8.5cm\epsfbox{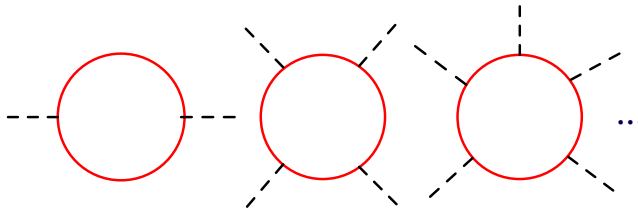}} 
\caption{The effective chiral lagrangian is the quark loop in the external chiral field,
or the determinant of the Dirac operator \ur{lagrangian}. Its real part is the
kinetic-energy term for pions, the Skyrme term and, generally, an infinite series 
in derivatives of the chiral field. Its imaginary part is the Wess--Zumino--Witten term
(with the correct coefficient), plus also an infinite series in derivatives~\cite{DP-CQSM}.} 
\end{figure}

An interesting question is, how does the effective lagrangian \ur{lagrangian} 
``know'' about the confinement of color? One writes \Eq{lagrangian} from the general
chiral symmetry considerations, and only the formfactors {\it e.g.} the dynamical mass 
$M(p)$ are subject to dynamical details. The difference between a confining and
a non-confining theory is hidden in the subtleties of the analytical behavior of
$M(p)$ and possible other formfactor functions in the Minkowski domain of momenta.
Specifically, the instanton model of the spontaneous chiral symmetry breaking~\cite{DP86} 
leads to such $M(p)$ that there is no real solution of the mass-shell equation $p^2=M^2(-p^2)$, 
meaning that quarks cannot be observable, only their bound states! However, 
this is not the only confinement requirement. Unfortunately, the instanton model's 
$M(p)$ has a cut at $p^2=0$ corresponding to massless gluons left in the model. 
In the true confining theory there should be no such cuts. 

In the bound states problems, however, quarks' momenta are space-like. Therefore,
one can use any reasonable falling function $M(p)$ reproducing the phenomenological
value of $F_\pi$ constant and of the chiral condensate~\cite{DP-CQSM}. As a matter
of fact, instantons do it phenomenologically very satisfactory. 

Constituent $u,d,s$ quarks necessarily have to interact with the $\pi,K,\eta$ 
fields according to \Eq{lagrangian}, and the dimensionless coupling constant is 
actually very large: $g_{\pi qq}(0)=\frac{M(0)}{F_\pi}\simeq 4$, where
the constituent quark mass $M(0)\simeq 350\,{\rm MeV}$ and $F_\pi\simeq 93\,{\rm MeV}$
are used.  

\begin{figure}[b]
\centerline{
\epsfxsize=8.5cm\epsfbox{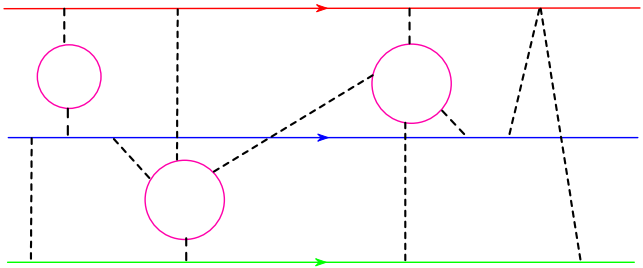}} 
\caption{Quarks in the nucleon (solid lines), interacting via pion
fields (dash lines).}
\end{figure}

The chiral interactions of constituent quarks in baryons, following from  
\Eq{lagrangian}, are schematically shown in Fig.~2. 
Antiquarks are necessarily present in the nucleon
as pions propagate through quark loops. The non-linear effects in
the pion field are essential since the coupling is strong. I
would like to stress that this picture is a model-independent
consequence of the spontaneous chiral symmetry breaking. 
One cannot say that quarks get a constituent mass but throw away
their strong interaction with the pion field. In principle,
one has to add perturbative gluon exchange on top of Fig.~2. However,
$\alpha_s$ is never really strong, such that gluon exchange can
be disregarded in the first approximation. The large value of
the pion-quark coupling suggests that Fig.~2 may well represent 
the most essential forces inside baryons. No ``confining strings''
are expected in the real world where it is energetically favorable
to break an expanded string by creating light pions. 

Although the low-momenta effective theory \ur{lagrangian} is a great simplification 
as compared to the microscopic QCD, as it uses the right degrees of freedom
appropriate at low energies, it is still a strong-coupling relativistic 
quantum field theory. Summing up all interactions inside the nucleon of the kind shown
in Fig.~2 is a difficult task. Maybe some day it will be solved numerically,
{\it e.g.} by methods presented by John Hiller in these Proceedings~\cite{Hiller}. 
In the meanwhile, it can be solved exactly in the limit of large number of colors $N_c$.  
With $N_c$ colors, the number of constituent quarks in a baryon is $N_c$, 
and all quark loop contributions in Fig.~2 are also proportional to $N_c$. 
Therefore at large $N_c$, quarks inside the nucleon create a large, nearly 
classical pion field: quantum fluctuations about the mean field are 
suppressed as $1/N_c$. The same field binds the quarks; therefore it is 
called the {\em self-consistent} field. [A similar idea is exploited in the 
shell model for nuclei and in the Thomas--Fermi approximation to atoms.] 
The problem of summing up all diagrams of the type shown in Fig.~2 is 
reduced to finding a classical self-consistent pion field. As long as $1/N_c$ 
corrections to the mean field results are under control, one can use the 
large-$N_c$ logic and put $N_c$ to its real-world value 3 at the end of 
the calculations. 

\begin{figure}[]
\centerline{\epsfxsize=9cm\epsfbox{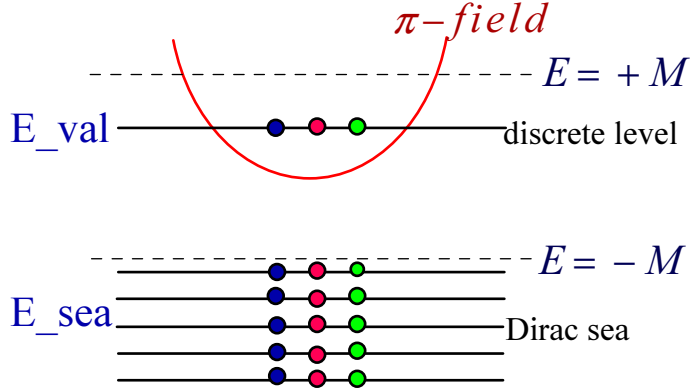}} 
\caption{If the trial pion field is large enough (shown schematically by
the solid curve), there is a discrete bound-state level for three `valence'
quarks, $E_{\rm val}$. One has also to fill in the negative-energy
Dirac sea of quarks (in the absence of the trial pion field it corresponds
to the vacuum). The continuous spectrum of the negative-energy levels is
shifted in the trial pion field, its aggregate energy, as compared to the
free case, being $E_{\rm sea}$. The nucleon mass is the sum
of the `valence' and `sea' energies, multiplied by three
colors, $M_N=3\left(E_{\rm val}[\pi(x)]+E_{\rm sea}[\pi(x)]\right)$.
The self-consistent pion field binding quarks is the one minimizing the nucleon
mass.}
\end{figure}

The model of baryons based on these approximations has been
named the Chiral Quark Soliton Model (CQSM)~\cite{DP-CQSM}. The
``soliton'' is another word for the self-consistent pion field in
the nucleon. However, the model operates with explicit
quark degrees of freedom, which enables one to compute any type of
observables, {\it e.g.} relativistic quark (and antiquark!)
distributions inside nucleons~\cite{SF}, and the quark light-cone
wave functions~\cite{light-cone}. In contrast to the naive quark models,
the CQSM is relativistic-invariant. Being such, it necessarily
incorporates quark-antiquark admixtures to the nucleon. Quark-antiquark
pairs appear in the nucleon on top of the three valence quarks
either as particle-hole excitations of the Dirac sea (read: mesons) or as
collective excitations of the mean chiral field.\\

There are two instructive limiting cases in the CQSM: \\ 
\vskip -0.2true cm

1. Weak $\pi(x)$ field. In this case the Dirac sea is weakly distorted as
compared to the no-field and thus carries small energy, $E_{\rm sea}\simeq 0$.
Few antiquarks. The valence-quark level is shallow and hence the three 
valence quarks are non-relativistic. In this limit the CQSM becomes very similar
to the constituent quark model remaining, however, relativistic-invariant and well
defined. \\
\vskip -0.2true cm

2. Large $\pi(x)$ field. In this case the bound-state level with valence
quarks is so deep that it joins the Dirac sea. The whole nucleon mass
is given by $E_{\rm sea}$ which in its turn can be expanded in the derivatives
of the mean field, the first terms being close to the Skyrme lagrangian.
Therefore, in the limit of large and broad pion field, the model formally reduces
to the Skyrme model. \\ 
\vskip -0.2true cm

{\em The truth is in between these two limiting cases.} The self-consistent pion
field in the nucleon turns out to be strong enough to produce a deep
relativistic bound state for valence quarks and a sufficient number of antiquarks,
so that the departure from the non-relativistic constituent quark model
is considerable. At the same time the self-consistent pion field is spatially
not broad enough to justify the use of the Skyrme model which is just a crude
approximation to the reality, although shares with reality some qualitative features. 
The CQSM demystifies the main paradox of the Skyrme model: how can one make a fermion
out of a boson-field soliton. Since the ``soliton'' is nothing but the self-consistent 
pion field that binds quarks, the baryon and fermion number of the whole construction
is equal to the number of quarks one puts on the valence level created by that field:
it is three in the real world with three colors.

\section{Baryon excitations}

There are excitations related to the fluctuations of the chiral field about its
mean value in the baryons. In the context of the Skyrme model many resonances
were found and identified with the existing ones in Ref.~\cite{HEHW,MK} and
quite recently in Ref.~\cite{Klebanov}. As I said before, the Skyrme model
is too crude, and one expects only qualitative agreement with the Particle Data.
The same work has to be repeated in the CQSM but it has not been done so far. 

There are also low-lying collective excitations related to slow rotation of the
self-consistent chiral field as a whole both in ordinary and flavor
spaces. The result of the quantization of such rotations was first given by
Witten~\cite{Witten}. The following $SU(3)$ multiplets arise as 
rotational states of a chiral soliton: $\left({\bf 8},\frac{1}{2}^+\right), 
\left({\bf 10},\frac{3}{2}^+\right), \left({\bf \overline{10}},\frac{1}{2}^+\right),
\left({\bf 27},\frac{3}{2}^+\right), \left({\bf 27},\frac{1}{2}^+\right)...$
They are ordered by increasing mass, see Fig.~4. The first two (the octet and the decuplet) 
are indeed the lowest baryons states in nature. They are also the only two
that can be composed of three quarks. However, the
fact that one can manage to obtain the correct quantum numbers of the octet 
and the decuplet combining only three quarks, does not mean that
they {\em are} made of three quarks only. The difficulties of such an interpretation
have been mentioned in the beginning. 

\begin{figure}[]
\vskip -0.8true cm
\centerline{\epsfxsize=8cm\epsfbox{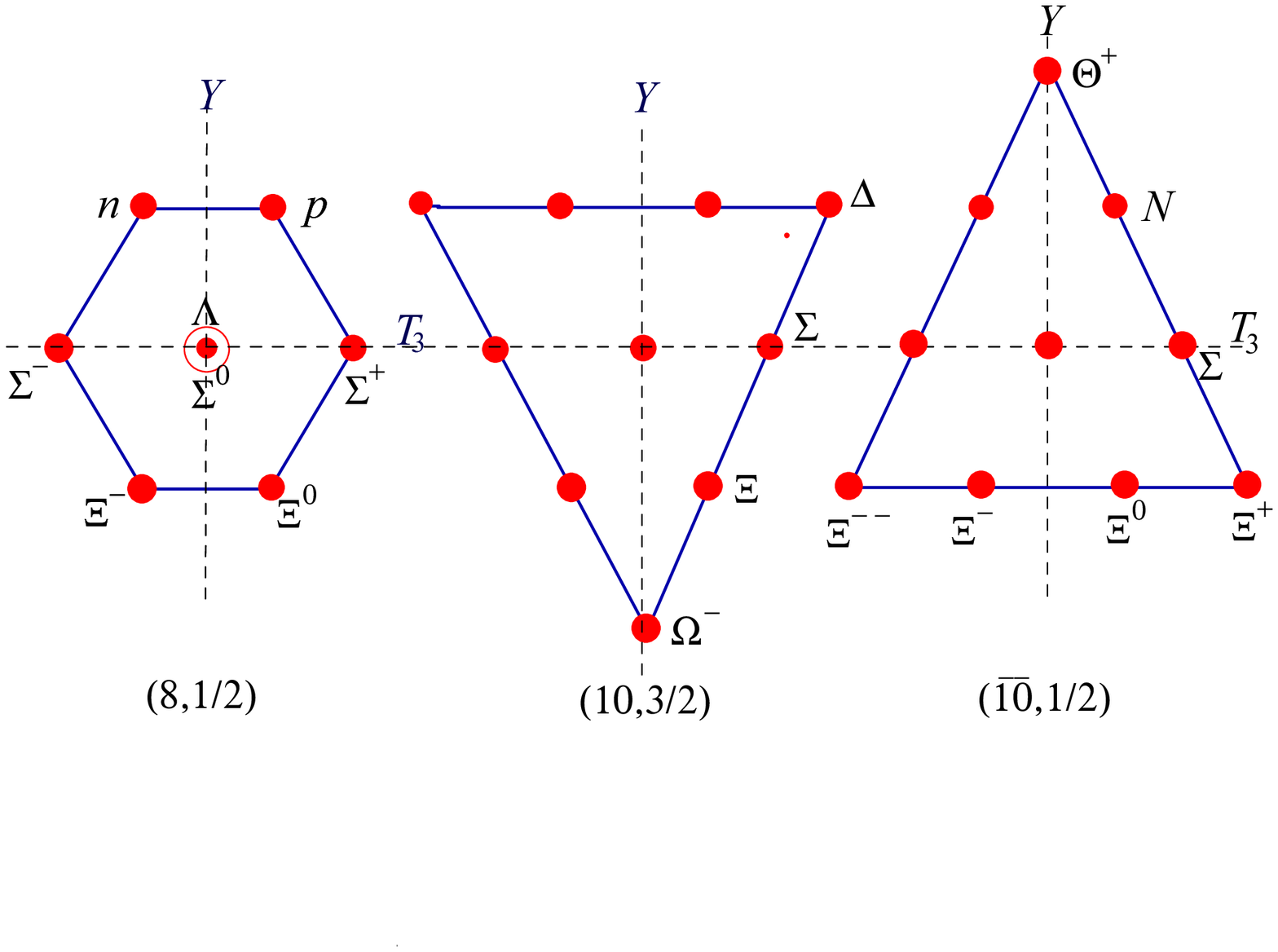}} 
\vskip -1.3true cm

\caption{The lowest baryon multiplets which can be interpreted
as rotational states in ordinary and 3-flavor spaces, shown in the $Y-T_3$
axes.}
\end{figure}

Therefore, one should not be {\em a priori} confused by the fact that higher-lying 
multiplets cannot be made of three quarks: even the lowest ones are not. A more
important question is where to stop in this list of multiplets. Apparently
for sufficiently high rotational states the rotations become too fast: 
the centrifugal forces will rip the baryon apart. Also the radiation of pions
and kaons by a fast-rotating body is so strong that the widths of
the corresponding resonances blow up~\cite{Regge}. Which precisely rotational excitation
is the last to be observed in nature, is a quantitative question:
one needs to compute their widths in order to make a judgement. If the width
turns out to be in the hundreds of MeV, one can say that this is where the rotational 
sequence ceases to exist. 

An estimate of the width of the lightest member of the antidecuplet, shown at the top 
of the right diagram in Fig.~4, the $\Theta^+$, gave a surprisingly small result:
$\Gamma_{\Theta}<15\,{\rm MeV}$~\cite{97}. This result obtained in the CQSM, 
immediately gave credibility to the existence of the antidecuplet. It should be 
stressed that there is no way to obtain a small width in the oversimplified Skyrme model.  

In pentaquarks forming the antidecuplet shown on the right of Fig.~4, the additional 
$Q\bar Q$ pair is added in the form of the excitation of the (nearly massless) 
chiral field. Energy penalty would be zero, had not the chiral field been restricted 
to the baryon volume. Important, the antidecuplet-octet splitting is not twice 
the constituent mass $2M$ but less. In the case of a large-size baryon it costs 
a vanishing energy to excite the antidecuplet~\cite{DP_Nc03}.

\section{Quark wave functions}

The wave function of baryons in the CQSM has been derived recently by Petrov and 
Polyakov~\cite{light-cone} in the infinite momentum frame. Here I translate it to the 
baryon rest frame. We shall see how easily one can get the non-relativistic
$SU(6)$ wave functions for ordinary octet and decuplet baryons ``from the soliton''.
Next, I derive the new result for the antidecuplet 5-quark wave functions. 

Let $a,a^\dagger({\bf p})$ and $b,b^\dagger({\bf p})$ be the annihilation--creation
operators of quarks and antiquarks (respectively) satisfying the usual 
anticommutator algebra. The vacuum $|\Omega_0\!>$ is such that $a,b|\Omega_0\!>=\!0$. 
According to the CQSM, a baryon is $N_c$ ``valence'' quarks on a discrete level
created by the self-consistent pion field, {\it plus} the negative-energy Dirac
sea of quarks, distorted as compared to the free case by the same self-consistent
pion field, see Fig.~2. At large $N_c$, the Dirac sea is given by the coherent
exponent
\beq
{\rm coherent\;exponent}=\exp\left(\int\!(d\zh{p})(d\zh{p'})\,a^\dagger(\zh{p})
W(\zh{p},\zh{p'})b^\dagger(\zh{p'})\right)|\Omega_0\!>, 
\la{cohexp}\eeq   
where $(d\zh{p})=d^3\zh{p}/(2\pi)^3$ and $W(\zh{p_1},\zh{p_2})$
is the finite-time quark Green function at equal times in the static
external field of the chiral ``soliton'', to be specified below. 
The valence quark part of the wave function is given by a product
of $N_c$ quark creation operators that fill in the discrete level:
\beqa\la{val1}
&&{\rm valence}=\prod_{{\rm color}=1}^{N_c}\int(d\zh{p})\,F(\zh{p})\,a^\dagger(\zh{p}),\\    
&&F(\zh{p})\!=\!\!\int\!(d\zh{p'})\!\left[u^*(\zh{p})f_{\rm lev}(\zh{p})(2\pi)^3
\delta(\zh{p}\!-\!\zh{p'})\!-\!W(\zh{p},\zh{p'})v^*(\zh{p'})f_{\rm lev}(-\!\zh{p})\right]\!,
\la{val2}\eeqa
where $f_{\rm lev}(\zh{p})$ is the Fourier transform of the wave function of the
level. The second term in \Eq{val2} is the contribution of the distorted Dirac sea
to the one-quark wave function; I shall neglect it for simplicity in what follows. 
With the same accuracy, the discrete level's wave function can be approximated by the 
upper component (as if it was non-relativistic):
\beq
F^{ij}(\zh{p})=\int\!(d\zh{p})e^{i\zh{p\cdot x}}\,\epsilon^{ij}\,h(r)
\la{F}\eeq
where $h(r)$ is the $L\!=\!0$ solution of the bound-state Dirac equation with energy 
$E\in[-M,M]$ for the given profile function of the soliton $P(r)$~\cite{DP-CQSM}:
\beqa\nn
&&h'=-M\sin P\,h+(E+M\cos P)\,j\,,\\
\nn
&&j'+\frac{2}{r}j=(M\cos P-E)\,h+M\sin P\,j\,.
\eeqa 
In the non-relativistic limit the $L\!=\!1$ function $j(r)$ is neglected in \Eq{F}. 
In \Eq{F} $i=1,2$ are spin and $j=1,2$ are isospin indices; $\epsilon^{ij}$
is the antisymmetric tensor. 

The $Q\bar Q$ pair wave function $W(\zh{p_1},\zh{p_2})$ determines the structure
of the Dirac continuum; it is also a matrix in both spin and isospin indices.
I denote by $(i,j)$ those of the quark and by $(i',j')$ those of the antiquark.
We shall need the Fourier transforms of all odd ($\Pi$) and all even ($\Sigma$) powers
of the self-consistent pion field:
\beqa\la{Pi}
&&\Pi^{j}_{j'}(\zh{q})=\int\!d\zh{r}\,e^{-i(\zh{q\cdot r})}\,
\left(\zh{n\cdot\tau}\right)^{j}_{j'}\,\sin P(r)\,,\\
\la{Si}
&&\Sigma^{j}_{j'}({\bf q})=\int\!d{\bf r}\,e^{-i(\zh{q\cdot r})}\,
\delta^{j}_{j'}\,(\cos P(r) -1)\,.
\eeqa  
Correspondingly, $W\!=W^{(\Pi)}+W^{(\Sigma)}$ can be divided into two pieces,
\beq
W^{ji\,(\Pi,\Sigma)}_{j'i'}({\bf p},{\bf p'})
=w^{i\,(\Pi,\Sigma)}_{i'}({\bf p},{\bf p'})\,
\Pi(\Sigma)^{j}_{j'}({\bf p+p'}),
\la{W}\eeq
where
\beqa\la{w}
w^{i\,(\Pi,\Sigma)}_{i'}&=&\frac{1}{2(\ee+\ee')}
\sqrt{\frac{MM'}{\ee\ee'(M+\ee)(M'+\ee')}}\\
\nn
&&\cdot\left\{\begin{array}{c}
\left[({\bf p\!\cdot\!p'})-(M+\ee)(M'+\ee')\right]\delta^{i}_{i'}\!
+i\ee_{pqr}p_{p}p'_{q}(\sigma_r)^{i}_{i'}\,,\\
\left[(M+\ee)p'_{r}-(M'+\ee')p_{r}\right](\sigma_r)^{i}_{i'}\,,
\end{array}\right.
\eeqa
with $\ee=\sqrt{M^2(\zh{p})+\zh{p}^2}$, the primed variables being 
related to the antiquark. In the coordinate space 
the pair wave function is given by a convolution of the self-consistent 
chiral field and the Fourier transforms of $w^{(\Pi,\Sigma)}$:
\beqa
\nn
&&W^{ji\,(\Pi,\Sigma)}_{j'i'}({\bf r,r'})\!
=\!\!\int\!d^3{\bf r''}\,w^{i\,(\Pi,\Sigma)}_{i'}({\bf r\!-\!r''},{\bf r'\!-\!r''})
\cdot \left\{\!\begin{array}{c}\left({\bf r''\cdot\tau}\right)^{j}_{j'}\,\sin P(r'')/r'',\\
\delta^{j}_{j'}\,(\cos P(r'')-1),\end{array}\right.\\
\nn\\
\la{wPiSi}
&&w^{i\,(\Pi,\Sigma)}_{i'}({\bf r\!-\!r''},{\bf r'\!-\!r''})\!
=\!\!\int\!\!(d\zh{p})(d\zh{p'})\,e^{i({\bf p\cdot
r\!-\!r''})}\,e^{i({\bf p'\cdot r'\!-\!r''})}\,w^{i\,(\Pi,\Sigma)}_{i'}({\bf p},{\bf p'}).
\eeqa
These functions can be computed numerically once the profile function of the 
self-consistent chiral field is known. \Eqs{w}{wPiSi} give the amplitudes 
of various spin, isospin and orbital $Q\bar Q$ states inside a baryon. 
The partial waves depend on the $Q\bar Q$ coordinates $(\zh{r,r'})$ with 
respect to the baryon center of mass. 

To get quark wave functions inside a particular baryon, one has to rotate all the 
isospin indices $j$'s, both in the discrete level and in the $Q\bar Q$ pairs, 
by an $SU(3)$ matrix $R^f_j,\;f=1,2,3,\;j=1,2$, and to project it to the specific
baryon from the $\left({\bf 8},\frac{1}{2}^+\right), \left({\bf 10},\frac{3}{2}^+\right)$
or $\left({\bf \overline{10}},\frac{1}{2}^+\right)$. ``Project'' means integrating
over the $SU(3)$ rotation matrices $R$ with a Haar measure normalized to unity. 
In full glory, the quark wave function inside a particular baryon $B$ with spin 
projection $k$ is given by 
\beqa\nn
&&\Psi^B_k=\int \!dR D^{B\,*}_k(R) \ee^{\alpha_1...\alpha_{N_c}}\prod_{n=1}^{N_c}
\int\!(d\zh{p_n})\,R^{f_n}_{j_n}F^{i_nj_n}(\zh{p_n})\,
a^\dagger_{\alpha_nf_ni_n}(\zh{p_n})\\
\la{Psi}
&&\cdot\exp\left(\int\!(d\zh{p})(d\zh{p'})\,a^\dagger_{\alpha f i}(\zh{p})R^f_j
W^{ji}_{j'i'}(\zh{p},\zh{p'})R^{\dagger\,j'}_{f'}b^{\dagger\,\alpha f'i'}(\zh{p'})\right)|\Omega_0\!>\,.
\eeqa  
Here $\alpha$ stands for color, $f$ for flavor and $i$ for spin indices. Let me 
give a few examples of the baryons' (conjugate) rotational wave functions $D^{B*}(R)$:
\beqa\la{n} 
&&{\rm neutron,\;spin\;projection}\;k:\qquad D^{n\,*}_k=\sqrt{8}\,\ee_{kl}R^{\dagger\,l}_2R^3_3,\\
\la{Deltappuu} 
&&\Delta^{++},\;{\rm spin\;projection}\;+\!
\frac{3}{2}:\qquad D^{\Delta^{++}\,*}_{\uparrow\uparrow}=
\sqrt{10}\,R^{\dagger\,2}_1R^{\dagger\,2}_1R^{\dagger\,2}_1,\\
\la{Delta0u} 
&&\Delta^{0},\,{\rm spin\;projection}\,+\!
\frac{1}{2}:\quad D^{\Delta^{0}\,*}_{\uparrow}\!=\!
\sqrt{10}R^{\dagger\,2}_2(2R^{\dagger\,2}_1R^{\dagger\,1}_2\!+\!R^{\dagger\,2}_2R^{\dagger\,1}_1),\\
\la{Theta}
&&\Theta^+,\;{\rm spin\;projection}\;k:\qquad D^{\Theta\,*}_{k}=\sqrt{30}\,R^3_3R^3_3R^3_k.
\eeqa

If the coherent exponent with $Q\bar Q$ pairs is ignored, one gets from the general \Eq{Psi}
the 3-quark Fock component of the octet and decuplet baryons. It depends on the quark ``coordinates'':
the position in space ($\zh{r}$), the color ($\alpha$), the flavor ($f$) and the spin ($i$), 
and also on the baryon spin $k$. For example, the neutron 3-quark wave function turns out to be
\beqa\nn
&&\left(|n\!>_k\right)^{f_1f_2f_3,i_1i_2i_3}({\bf r_1,r_2,r_3})
=\ee^{f_1f_2}\,\ee^{i_1i_2}\,\delta^{f_3}_2\,\delta^{i_3}_k\,
h(r_1)h(r_2)h(r_3)\\
&&+\,{\rm permutations\;of\;1,2,3},
\la{n1}\eeqa
antisymmetrized in color. It is better known in the form
\beqa\nn
|n\!\uparrow>&\!=\!&2\,d\!\uparrow\!(r_1)d\!\uparrow\!(r_2)
u\!\downarrow\!(r_3)\!-\!d\!\uparrow\!(r_1)u\!\uparrow\!(r_2)
d\!\downarrow\!(r_3)\!-\!u\!\uparrow\!(r_1)d\!\downarrow\!(r_2)
d\!\uparrow\!(r_3)\\
&+& {\rm permutations\;of\;} r_1,r_2,r_3,
\la{n2}\eeqa
which is the well-known non-relativistic $SU(6)$ wave function of the nucleon!
Petrov and Polyakov~\cite{light-cone} have obtained the corresponding $SU(6)$ function
in the infinite-momentum frame. 

Performing the group integration with the {\it decuplet} rotational functions 
\urs{Deltappuu}{Delta0u} one also gets the well-known $SU(6)$ wave functions 
in the non-relativistic limit. Relativistic corrections to those wave functions 
are easily computable from \Eq{val2}, as are the 5-quark Fock components of the 
usual octet and decuplet baryons. To find those, one needs to expand the coherent 
exponent in \Eq{Psi} to the linear order in the additional $Q\bar Q$ pair, 
and perform the $SU(3)$ projecting. The result will be given in a subsequent publication. 
Here I shall go straight to the $\Theta^+$.

Projecting the three quarks from the discreet level on the $\Theta$ rotational 
function \ur{Theta} gives an identical zero, in accordance with the fact that
the $\Theta$ cannot be made of 3 quarks. The non-zero projection is achieved
when one expands the coherent exponent to the linear order. One gets then the 
5-quark component of the $\Theta$ wave function: 
\beqa\nn
&&|\Theta^+_k\!>^{f_1f_2f_3f_4,i_1i_2i_3i_4}_{f_5,i_5}({\bf r_1\ldots r_5})
=\ee^{f_1f_2}\ee^{f_3f_4}\delta^3_{f_5}\,\ee^{i_1i_2}\\
&&\cdot\, h(r_1)h(r_2)h(r_3)\,
W^{i_3i_4}_{k\,i_5}({\bf r_4,r_5})+{\rm permutations\;of\;1,2,3}.
\la{Theta1}\eeqa
The color structure of the antidecuplet wave function is
$\ee^{\alpha_1\alpha_2\alpha_3}\delta^{\alpha_4}_{\alpha_5}$.
Indices 1 to 4 refer to quarks and index 5 refers to the antiquark,
in this case $\bar s$ thanks to $\delta^3_{f_5}$. The quark
flavor indices are $f_{1\!-\!4}=1,2=u,d$. Naturally, we have obtained
$\Theta^+=uudd\bar s$. 

We see that the first two valence $u,d$ quarks from the discrete level form a spin- 
and isospin-singlet diquark (although not correlated in space), like in the
nucleon, see \Eq{n1}. However, the other pair of quarks do not form a 
similar spin-zero diquark. For example, in the ``$\Sigma$'' part of the wave 
function the $\Theta$ spin $k$ is determined by the spin of the third quark 
from the discrete level. Since in the CQSM the functions $h(r_{1,2,3})$ and 
$W(\zh{r_4},\zh{r_5})$ are known, \Eq{Theta1} gives the complete
color, flavor, spin and space 5-quark wave function of the $\Theta^+$ in its rest
frame. The 5-quark wave functions of other members of the antidecuplet
can be obtained in a similar manner. 

For the computation of the $\Theta$ width, this wave function is, however, 
inadequate as a matter of principle. As explained in Refs.~\cite{DP_mixing,D04}, 
the only consistent way to compute the width is using the infinite momentum 
frame~\cite{light-cone} where there is no pair creation or annihilation, 
and the Fock decomposition is well defined. In that frame, the decay of the 
$\Theta^+$ goes into the {\em five}-quark component of the nucleon only. 
It is first of all suppressed to the extent the 5-quark component of the 
nucleon is less than its 3-quark component. An additional suppression comes 
from the spin-flavor overlap. A preliminary crude estimate shows that the 
$\Theta^+$ width can be extremely small. \\

\noindent{\bf Acknowledgments}\\
\vspace{-0.3cm}

I thank the organizers of the {\it Continuous Advances} for hospitality,
and V.~Petrov and M.~Polyakov for numerous discussions. This
work has been supported in part by the DOE under contract DE-AC05-84ER40150.

\end{document}